\documentclass[preprint,superscriptaddress,amsmath,amssymb,longbibliography]{revtex4-1}  
\usepackage{graphicx}
\usepackage{bm}
\usepackage[colorlinks=true,urlcolor=blue,citecolor=blue,linkcolor=blue,bookmarks=false,pdfstartview={FitH}]{hyperref}      
\usepackage{color}
\usepackage{multirow}

\def \cm-1{cm$^{-1}$\,}

\begin{document} 
\title{Direct observation of phonon mode with an antisymmetric Raman tensor}   
                         
\author{Shangfei Wu$^\star$}
\email{wusf@baqis.ac.cn}
\affiliation{Beijing Academy of Quantum Information Sciences, Beijing 100193, China}
\affiliation{Department of Physics and Astronomy, Rutgers University, Piscataway, New Jersey 08854, USA}
\author{Shang~Ren$^\star$}
\affiliation{Department of Physics and Astronomy, Rutgers University, Piscataway, New Jersey 08854, USA}
\author{Kai~Du}
\affiliation{Department of Physics and Astronomy, Rutgers University, Piscataway, New Jersey 08854, USA}
\affiliation{Keck Center for Quantum Magnetism, Rutgers University, Piscataway, New Jersey 08854, USA}
\author{Xianghan~Xu}
\affiliation{Department of Physics and Astronomy, Rutgers University, Piscataway, New Jersey 08854, USA}
\affiliation{Keck Center for Quantum Magnetism, Rutgers University, Piscataway, New Jersey 08854, USA}
\author{Sang-Wook~Cheong} 
\affiliation{Department of Physics and Astronomy, Rutgers University, Piscataway, New Jersey 08854, USA}
\affiliation{Keck Center for Quantum Magnetism, Rutgers University, Piscataway, New Jersey 08854, USA}
\author{David~Vanderbilt} 
\affiliation{Department of Physics and Astronomy, Rutgers University,
Piscataway, New Jersey 08854, USA}
\author{Girsh~Blumberg} 
\email{girsh@physics.rutgers.edu}
\affiliation{Department of Physics and Astronomy, Rutgers University, Piscataway, New Jersey 08854, USA}
\affiliation{National Institute of Chemical Physics and Biophysics,
12618 Tallinn, Estonia}
\date{\today}           

\maketitle

\[ \textbf{Abstract} \]
                                                    
Phonon angular momentum, characterized by circular or elliptical motion of atoms, plays a decisive role in diverse phenomena, ranging from the phonon Hall effect, phonon magnetic moment, Einstein-de Haas effect, Weyl phonons, and driven chiral phonons. \textcolor{black}{The phonon modes with an antisymmetric Raman tensor, which carry an intrinsic phonon angular momentum in a pseudovector symmetry channel, have been overlooked in the past in the chiral phonon community.}
Here, we report the detection of this type of phonon mode using polarization-resolved Raman spectroscopy. We observe a phonon at 266\,\cm-1 in the antiferromagnetic chiral phase of Cr$_2$O$_3$ \textcolor{black}{in the $A_2$-pseudovector symmetry channel}. 
This $A_2$ mode's frequency and the elliptical motion are captured by first-principles phonon calculations, which treat both lattice and spins on an equal footing as `slow' degrees of freedom. \textcolor{black}{This mode gets Raman activity due to the presence of Cr vacancies in Cr$_2$O$_3$ that locally break the twofold rotational symmetry perpendicular to the threefold axis.}
Our results provide a new way to detect the novel symmetry-forbidden $A_2$ phonon modes with an antisymmetric Raman tensor in a pseudovector symmetry channel.

\newpage

\textbf{Introduction} 

Chiral phonons, characterized by circular or elliptical motion of ions, carry a nonzero phonon angular momentum (PAM)~\cite{McLellan_JPC_1988,Zhang_2014_PhysRevLett,Zhang_2015_PhysRevLett,Zhu_2018_Science,Ueda_2023_RIXS,Wang_2024_review,Zhang_arxiv_2025,yang_2025_catalog,Juraschek2025,Zhang_2026_PRB}. They play important roles in diverse phenomena of condensed matter, ranging from the phonon Hall effect~\cite{Grissonnanche_2020_NP,Zhang_2025_NP}, phonon magnetic moment~\cite{Juraschek_2017_PhysRevMaterials,Juraschek_2019_PhysRevMaterials,Ren_2021_PhysRevLett}, Einstein–de Haas effect~\cite{Tauchert_2022_Nature}, Weyl phonons~\cite{Zhang_NC_2025}, and driven chiral phonons~\cite{Nova_2017,Luo_2023_science,Zeng_2025_science,Romao_2024_acsnano}.
The phonon magnetic moment is typically on the order of the ionic magneton, which is on the order of magnitude about 10$^{-4} \mu_\text{B}$. Recent experiments have revealed that the phonon magnetic moments can be 2 to 4 orders of magnitude larger than the typical ionic magneton due to the coupling to the electronic degree of freedom~\cite{SCHAACK_1975,Ahrens_1979_PhysRevLett,Cheng_2020_NanoLetter,Baydin_2022_PhysRevLett,Lujan_2024_PNAS,Mustafa_2025_ACS_Nano,Che_2025_PhysRevLett,Yang_2025_PhysRevLett,Wu_2025_FMO_PRL,Chaudhary_2024_PhysRevB}. \textcolor{black}{These experiments identify right and left circularly polarized chiral phonons close to the Brillouin zone center, originating from the splitting of the doubly degenerate modes, either in an external magnetic field~\cite{Ahrens_1979_PhysRevLett,Cheng_2020_NanoLetter,Baydin_2022_PhysRevLett,Lujan_2024_PNAS,Mustafa_2025_ACS_Nano,wang_2025_Huang} or below an internal magnetization temperature~\cite{Che_2025_PhysRevLett,Yang_2025_PhysRevLett,Wu_2025_FMO_PRL,Bao_2026_2026}. However, the non-degenerate chiral modes with intrinsic nonzero PAM close to the Brillouin zone center---not originating from the splitting of the doubly degenerate modes---have been rarely explored.}
                   
The antisymmetric response, which experimentally exhibits $A_{2g}$ or $A_2$ symmetries, has attracted great interest in the study of the electronic and magnetic excitations in recent years~\cite{Koningstein1968,Shastry1990,Khveshchenko1994,Liu1993,Devereaux2007RMP,Kung2015,Riccardi2016,Kung2017PRL,Kung_2018_A2g,He_arxiv_2025,Hsu_2025_PhysRevB,shan2025_chiral_spin_mode,udina_2025_antisymmetric}.
These $A_{2g}$ or $A_2$ responses have the properties of the antisymmetric Raman tensor ($\alpha_{ij}$) with the form $\alpha_{ij}=-\alpha_{ji}$. 
For phononic excitations, the Raman crosssection is proportional to the square of the Raman tensor element, where the Raman tensor is constrained to a symmetric form $\alpha_{ij}=\alpha_{ji}$ in the quasistatic, adiabatic, or polarizability theory of Raman scattering for nonmagnetic media; thus, the phononic excitations with antisymmetric Raman tensor, \textcolor{black}{neither Raman nor infrared active in linear order}, are generally not detectable~\cite{Placzek1934,Loudon2012}. \textcolor{black}{Nevertheless, they could emerge when the polarizability description breaks down near resonance conditions when the frequency of the incident light is within an absorption band~\cite{Placzek1934,Loudon2012,Spiro_1972_PNAS,Cerdeira_1986_PhysRevLett,Guntherodt_1979_PhysRevB}.}
In magnetic systems like CrI$_3$, time-reversal symmetry breaking can induce an antisymmetric component in the Raman tensor of the fully symmetric phonon mode~\cite{Huang_2020_NN}. 
Theoretically, the basis functions of the antisymmetric $A_{2g}$ irreducible representation (e.g.,~$D_{3d}$ point group) transform as the $z$ component of the angular momentum $L_z$ (in the lowest order), which is a pseudovector~\cite{Ovander1960,Koster1963}. This type of phononic excitation links angular momentum and the circular motion.
A previous polarizability calculation based on a classical model of an electron rotating in a circular orbit in a centrosymmetric field finds that the polarizability contains purely imaginary antisymmetric terms, which change their sign when the rotation direction is reversed~\cite{Ovander1960,Baranova_1978_JournalRS}. When the electron is replaced with an ion in this model, a link is expected between PAM due to circular motion and the antisymmetric Raman response. Thus, a non-degenerate chiral phonon with nonzero PAM carried by $A_{2g}$ or $A_2$ phonons has yet to be explored.
                 
In this paper, we detect an $A_2$ phonon at 266\,\cm-1 in the antiferromagnetic (AFM) chiral phase of Cr$_2$O$_3$ using polarization-resolved Raman spectroscopy.  This $A_2$ mode's frequency and the elliptical motion are captured by first-principles phonon calculations, which treat both lattice and spins on an equal footing as `slow' degrees of freedom.
\textcolor{black}{The presence of Cr vacancies in Cr$_2$O$_3$ activates this silent $A_2$ mode by locally breaking the twofold rotational symmetry perpendicular to the threefold axis.}
Our results provide a new method to detect the novel $A_2$ phonons with an antisymmetric Raman tensor in a pseudovector symmetry channel.

\textbf{Results}\label{Results}

\textbf{Phonon modes.} 

Above $T_\text{N}=307.5$\,K, Cr$_2$O$_3$ belongs to a trigonal corundum nonmagnetic crystal structure with space group $R\bar{3}c$ (No. 167, point group $D_{3d}$). The Cr$^{3+}$ atoms are located at an octahedral center with six surrounding O$^{2-}$ atoms, resulting in a rhombohedral primitive cell with four Cr atoms along the trigonal $c$ axis (\textbf{Methods}).
From group theoretical considerations, \textcolor{black}{the $\Gamma$ point phonon modes of  Cr$_2$O$_3$ can be expressed as $\Gamma$ = 2$A_{1g}$ $\oplus$ 2$A_{1u}$ $\oplus$ 3$A_{2g}$ $\oplus$ 3$A_{2u}$ $\oplus$ 5$E_{u}$ $\oplus$ 5$E_{g}$.}
Raman-active modes are \textcolor{black}{$\Gamma_{\text{Raman}}$= 2$A_{1g}$ $\oplus$ 5$E_g$} and infrared-active modes are $\Gamma_{\text{IR}}$ = 2$A_{2u}$ $\oplus$ 4$E_{u}$.  
\textcolor{black}{Note that $A_{2g}$ is neither Raman nor infrared active in linear order.}
Below $T_\text{N}$, Cr$_2$O$_3$ forms a two-sublattice colinear AFM order with the easy axis along the $c$-axis direction. The magnetic space group is $R\bar{3}'c'$.
The presence of magnetic ordering in the bulk Cr$_2$O$_3$ breaks parity ($i$), dihedral mirror ($\sigma_d$), and rotoinversion ($S_6$) symmetries associated with the $D_{3d}$ point group. Consequently, the original $D_{3d}$ point group symmetry is reduced to the $D_3$ below $T_\text{N}$, resulting in interirrep mixing between bare phonons of different irreducible representations (irreps). 
The $E_g$ and $E_u$ irreps become the $E$ irrep, the $A_{1g}$ and $A_{1u}$ become the $A_1$ irrep, and the $A_{2g}$ and $A_{2u}$ irreps become the $A_2$ irrep of $D_3$ (Supplementary Note~\ref{GroupTheory}). 
\textcolor{black}{From group theoretical considerations, the optical modes are $\Gamma_{\text{optical}}$ = 4$A_1$ $\oplus$ 5$A_2$ $\oplus$ 9$E$, and the acoustic modes are $\Gamma_{\text{acoustic}}$ =$A_2$ $\oplus$ $E$. 
All the $E$-symmetry optical modes are both Raman and infrared active. The $A_1$-symmetry modes are Raman active, while the $A_2$-symmetry modes become infrared active.} 
In the following, we use the irreps of the $D_3$ point group to label the optical phonons.

\begin{figure*}[!t] 
\begin{center}
\includegraphics[width=\columnwidth]{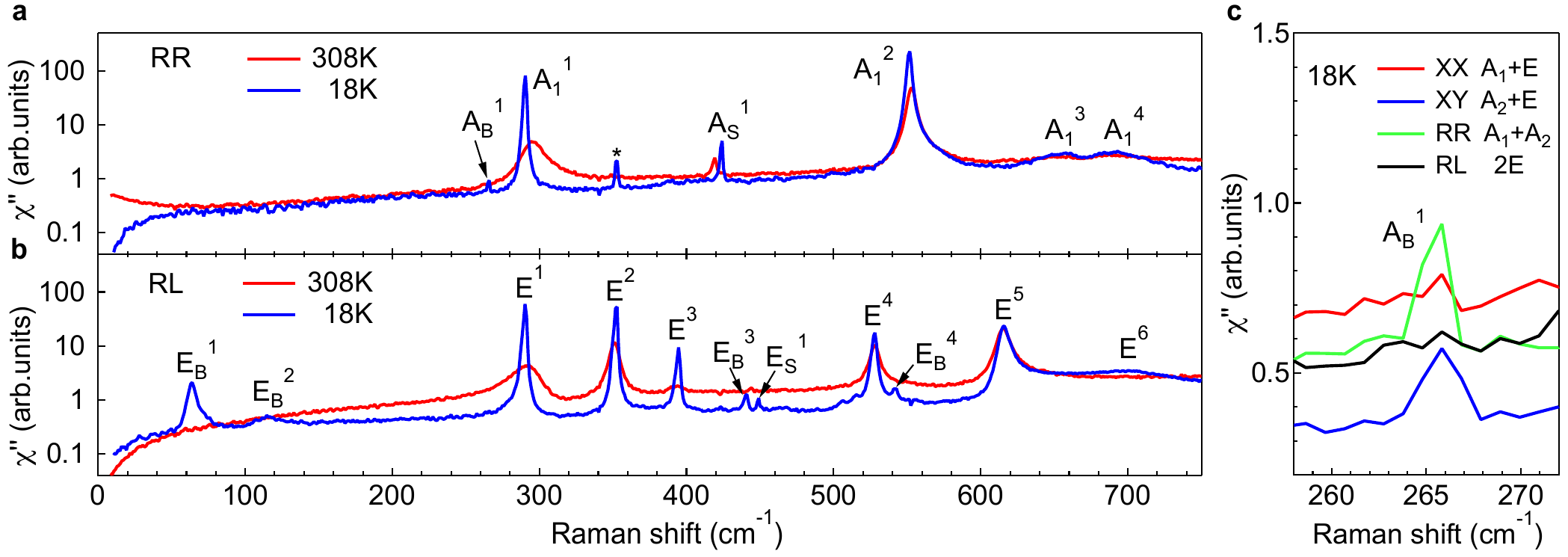}
\end{center}
\caption{\label{Fig1_RR_RL} 
\textbf{Phonon modes.}  Raman response of Cr$_2$O$_3$ at 308\,K and 18\,K for $RR$ [$A_1$ + $A_2$] (panel \textbf{a}) and $RL$ [$E$] (panel \textbf{b}) scattering geometries from the $ab$ surface with 647\,nm laser excitation. \textbf{c} Polarization dependence of the mode at 266\,\cm-1 in four scattering geometries. The star in (\textbf{a}) marks the leakage intensity for the $E^2$  at  350\,\cm-1.
}
\end{figure*}    

In Fig.~\ref{Fig1_RR_RL}a,b, we present the Raman response for Cr$_2$O$_3$ from the $ab$ plane at 308\,K and 18\,K in the $RR$ and $RL$ scattering geometries. Two $A_1$ modes ($A^1_1$ mode at 290\,\cm-1 and $A^2_1$ mode at 552\,\cm-1) are observed in the $RR$ scattering geometry shown in Fig.~\ref{Fig1_RR_RL}a. Five $E$ modes ($E^1$ mode at 290\,\cm-1, $E^2$ mode at 352\,\cm-1, $E^3$ mode at 395\,\cm-1, $E^4$ mode at 528\,\cm-1, $E^5$ mode at 615\,\cm-1) are observed in the $RL$ scattering geometry shown in Fig.~\ref{Fig1_RR_RL}b.
These are the Raman-active bulk phonon modes expected from the group theoretical analysis and DFT phonon calculations (Table~\ref{phonon}). These phonon peak positions are consistent with a previous Raman scattering study~\cite{Shim_2004_PhysRevB}.

\begin{figure*}[!t] 
\begin{center}
\includegraphics[width=\columnwidth]{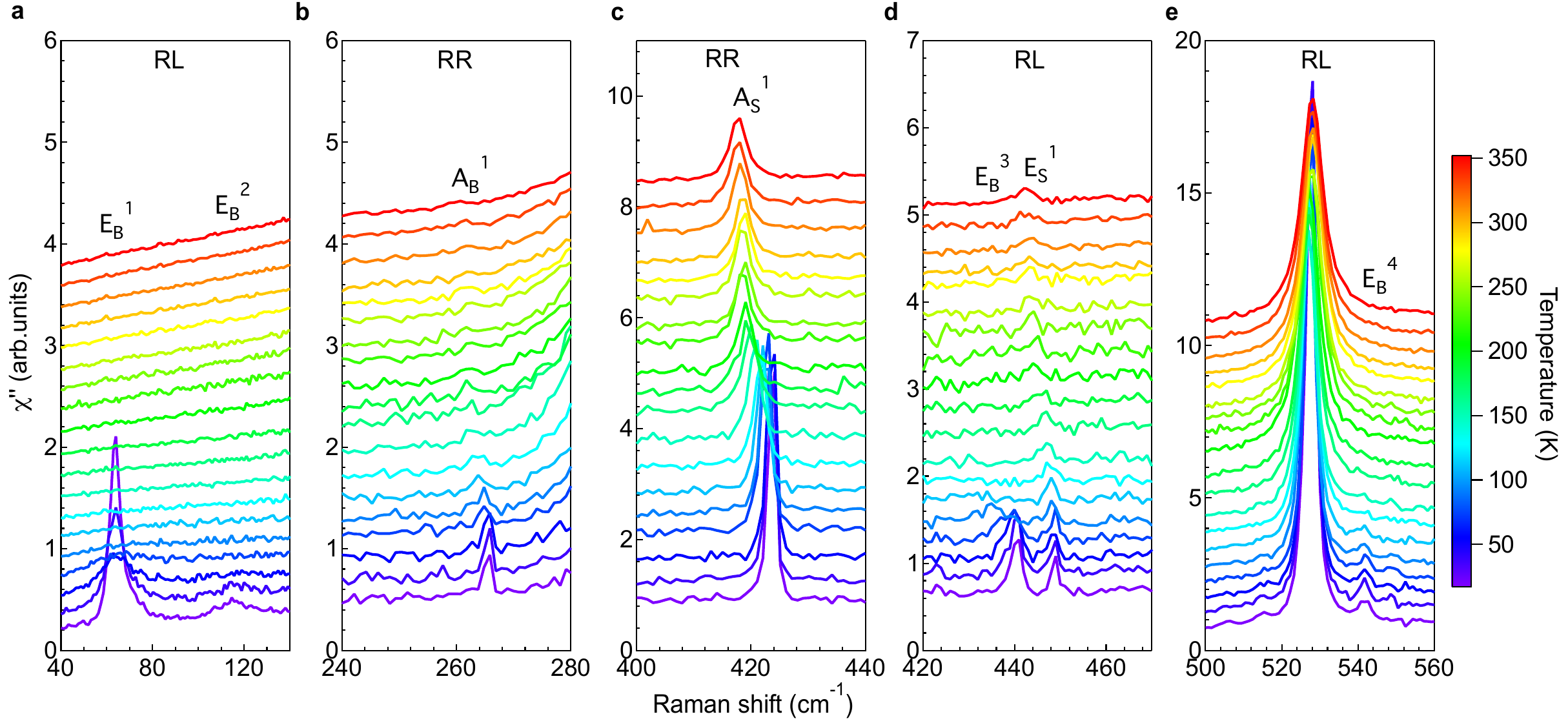}
\end{center}
\caption{\label{Fig2_T_dependence} 
\textbf{Temperature dependence.} Temperature dependence of the modes for the $E^1_\text{B}$ mode at 64\,\cm-1 and $E^2_\text{B}$ mode at 115\,\cm-1(panel \textbf{a}),
the $A^1_\text{B}$ mode at 266\,\cm-1(panel \textbf{b}), $A^1_\text{S}$ mode at 424\,\cm-1(panel \textbf{c}), $E^3_\text{B}$ mode at 440\,\cm-1 and $E^1_\text{S}$ mode at 449\,\cm-1(panel \textbf{d}), and $E^4_\text{B}$ mode at 542\,\cm-1(panel \textbf{e}).}
\end{figure*}

In addition to these bulk phonon modes, we detect two additional modes that are present at both 308\,K and 18\,K. One mode is located at 419\,\cm-1 at 300\,K, labeled as $A_\text{S}^1$. The peak position for this mode is close to the $A_{1u}$ mode predicted by DFT phonon calculations (Table~\ref{phonon}). The temperature dependence for this mode is shown in Fig.~\ref{Fig2_T_dependence}c. This mode shifts to 424\,\cm-1 at 18\,K.
Another mode is located at 444\,\cm-1 at 300\,K, labeled as $E_\text{S}^1$. The peak position for this mode is close to the $E_{u}$ mode predicted by DFT phonon calculations (Table~\ref{phonon}). The temperature dependence for this mode is shown in Fig.~\ref{Fig2_T_dependence}d. This mode shifts to 449\,\cm-1 at 18\,K.
According to the group theoretical analysis, the $A_{1u}$ and $E_{u}$ modes are infrared active \textcolor{black}{but Raman forbidden}. Since the $A_\text{S}^1$ and $E_\text{S}^1$ modes exist above and below $T_\text{N}$, we interpret these two modes as surface vibrational modes \textcolor{black}{following the wallpaper group symmetry $C_3$, because the inversion and mirror symmetries are broken at the crystal surface, assuming that half of the primitive unit cell is removed by polishing~\cite{Kung_2017_PhysRevB}.}

We then switch to the additional modes that appear only below $T_\text{N}$. The modes at 660\,\cm-1(labeled as $A^3_1$) and 690\,\cm-1 (labeled as $A^4_1$) in the $RR$ scattering geometry, and the mode at 700\,\cm-1 (labeled as $E^6$) in the $RL$ scattering geometry, only appear in the 18\,K data below $T_\text{N}$. These three modes’ energies are close to twice of the magnon energy at the Brillouin zone boundary~\cite{Samuelsen_1969_Solid}. Thus, these three modes are interpreted as two-magnon scattering peaks due to flipping a pair of neighboring spins in different sublattices below $T_\text{N}$~\cite{Girsh_PRB1994,Blumberg_1996_PhysRevB} \textcolor{black}{(references therein)}.

At low frequencies below 200\,\cm-1, two additional modes (one mode at 64\,\cm-1 labeled as $E_B^1$ and another one at 115\,\cm-1 labeled as $E_B^2$) appear in the 18\,K Raman data in the $RL$ scattering geometry.
\textcolor{black}{These modes are also confirmed at the same peak positions in the terahertz absorption spectra at 3\,K~\cite{Wu_arxiv2025_1}.} 
The temperature dependence for these two modes is shown in Fig.~\ref{Fig2_T_dependence}a. They appear only at low temperatures, roughly at about 100$\sim$150\,K. 
\textcolor{black}{
Compared with a cleaner sample, the Cr$_2$O$_3$ sample (the current study) shows a wider spin-flop transtion and 100 times stronger intensity in the terahertz absorption spectra for the $E_B^1$ mode~\cite{Wu_arxiv2025_1}.
}
Thus, these two modes are interpreted as defect-induced modes arising from a minute concentration of Cr vacancies in the bulk sample~\cite{Wu_arxiv2025_1}.

We notice that  the mode at 440\,\cm-1 (labeled as $E^3_\text{B}$), and the mode at 542\,\cm-1 (labeled as $E^4_\text{B}$) in the $RL$ scattering geometry appear in the 18\,K Raman data shown in Fig.~\ref{Fig1_RR_RL}a,b. The temperature dependence for the  $E^3_\text{B}$ and $E^4_\text{B}$ modes is shown in Fig.~\ref{Fig2_T_dependence}d,e, respectively. These modes appear below a temperature of about 100$\sim$150\,K, similar to the Cr defect modes $E_B^1$ and $E_B^2$.
The $E^3_\text{B}$ mode at 440\,\cm-1 and $E^4_\text{B}$ mode at 542\,\cm-1 correspond to the $E_u$ mode in the nonmagnetic phase ($E$ mode in the AFM phase) based on Table.~\ref{phonon}.
The $E_u$ modes are infrared active and Raman forbidden according to the group theoretical analysis. \textcolor{black}{The existence of these infrared-active $E_u$ modes in a Raman spectrum indicates interirrep mixing in the AFM phase of Cr$_2$O$_3$}. Since these modes ($E^3_\text{B}$ and $E^4_\text{B}$) appear at similar temperatures as the Cr defect modes $E_B^1$ and $E_B^2$, \textcolor{black}{we interpret the Raman activity of these two modes as mainly due to local twofold rotational symmetry (perpendicular to the threefold axis) breaking with the point group symmetry reduced to $C_3$ around the minute concentration of Cr vacancies.}

\begin{table*}[t]
\caption{\label{phonon} Experimental phonon frequencies at the Brillouin zone center for Cr$_2$O$_3$ at 18\,K and the phonon frequencies calculated by DFT. The units are in \cm-1. }
\begin{ruledtabular}
\begin{tabular}{ccccccc}
 Symmetry group $D_{3d}$ &  Symmetry group $D_3$ & DFT~\cite{LARBI_2017_Journal} &DFT~\cite{Fechner_2018_PhysRevMaterials}& DFT (current work) &Experiment (18\,K) \\
 \hline
 $A_{2g}$ & $A_2$   & 274&267 & 269& 266&  \\
 $A_{1g}$ & $A_1$  & 317 &310& 293& 290 &  \\
 $E_g$ & $E$   & 321&307& 294& 290 &  \\
 $E_u$ & $E$   & 336 &310& 308&-&  \\
 $E_g$ & $E$   & 369&357& 353 & 352& \\
 $E_g$ & $E$   & 400&414& 400 & 395&  \\
 $A_{2u}$ & $A_2$   & 424& 407& 408&-& \\
 $A_{1u}$ & $A$   & 432 &-& 417& 424(surface)&  \\
 $E_u$ & $E$   & 456 & 450&450& 440, 449(surface)& \\
 $A_{2g}$ & $A_2$   & 474 &460& 460&-&  \\
 $E_g$ & $E$  & 517&537& 525 & 528 &  \\
 $E_u$ & $E$  & 523&567& 544 & 542 &  \\
 $A_{2u}$ & $A_2$   & 527&574& 547 &-&  \\
 $A_{1g}$ & $A_1$  & 543 &577& 547 & 552&  \\
 $E_u$ & $E$  & 607 &634&  613&-&  \\
 $E_g$ & $E$   & 609&640& 618 & 615&  \\
 $A_{1u}$ & $A_1$   & 640&-& 623 &-&  \\
 $A_{2g}$ & $A_2$   & 693&690& 674 &-&  \\
\end{tabular}
\end{ruledtabular}
\end{table*}

\begin{figure*}[!t] 
\begin{center}
\includegraphics[width=0.8\columnwidth]{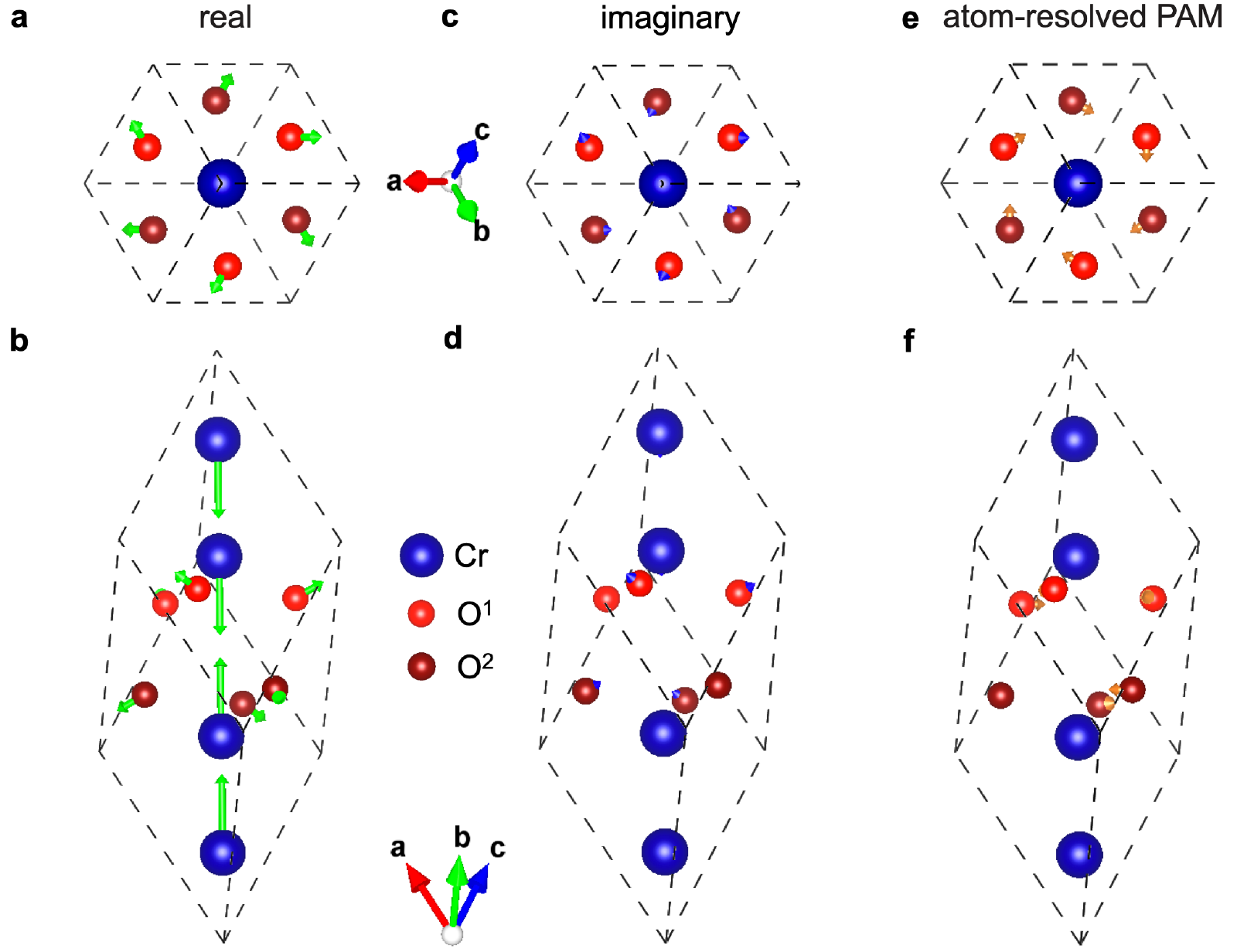}
\end{center}
\caption{\label{Fig3_motion} 
\textbf{DFT phonon calculations.} The real and imaginary components, as well as
atom-resolved PAM, of the antisymmetric $A_2$ phonon around 266\,\cm-1 from the DFT phonon calculations for
Cr$_2$O$_3$. The Cr atoms are shown in blue, while the top-layer
and bottom-layer O atoms are depicted in bright and dark red,
respectively. The green vectors indicate the real part of the
phonon displacement in top view (panel \textbf{a}) and side view (panel \textbf{b}).
The blue vectors indicate the imaginary
part of the phonon displacement in top view (panel \textbf{c}) and side view (panel \textbf{d}).
The brown vectors represent the atom-resolved PAM in top view (panel \textbf{e}) and side view (panel \textbf{f}).
The amplitude of atom-resolved phonon angular momentum in this figure is amplified by a factor of 5000 compared to that of $E$ mode shown in Fig.~6 of Ref.~\cite{Ren_2024_PhysRevX}.}
\end{figure*}      
                                                    
\textbf{$A_2$ mode and polarization dependence.}  

\textcolor{black}{We now turn to the mode at 266\,\cm-1(labeled as $A^1_\text{B}$) in the $RR$ scattering geometry, which is a main focus of this work.}
It appears in the 18\,K Raman data shown in Fig.~\ref{Fig1_RR_RL}a,b. The temperature dependence for the $A^1_\text{B}$ mode is shown in Fig.~\ref{Fig2_T_dependence}b. The mode appears below a temperature of about 100$\sim$150\,K, similar to the Cr defect modes $E_B^1$ and $E_B^2$.
The $A^1_\text{B}$ mode at 266\,\cm-1 corresponds to the $A_{2g}$ mode in the nonmagnetic phase ($A_2$ mode in the AFM phase) based on Table~\ref{phonon}. 

In Fig.~\ref{Fig1_RR_RL}c, we present the polarization dependence of the $A^1_\text{B}$ mode at 266\,\cm-1 in $XX$, $XY$, $RR$, and $RL$ scattering geometries at 18K. This mode appears in the $XY$ and $RR$, but not the $XX$ and $RL$ scattering geometries, supporting that it is an $A_2$ mode in the notations of the $D_3$ point group in the AFM phase. The basis function of the $A_2$ irrep transform as the $z$ component of the angular momentum $L_z$, which is a pseudovector~\cite{Ovander1960,Koster1963}. This suggests that the observed $A_2$ peak corresponds to a phonon mode with an antisymmetric Raman tensor, which is characterized by a circular or elliptical motion of atoms. 

\textbf{Phonon angular momentum.} 

To further demonstrate that the observed $A_2$ mode has circular or elliptical motions of vibration, we show the real and imaginary components of the atomic vibration patterns for the $A_2$ mode at 266\,\cm-1 in Fig.~\ref{Fig3_motion}a-d based on DFT phonon calculations.
To have a circular or elliptical motion of atoms, the real and imaginary components for the atomic vibration have to be noncollinear; otherwise, they combine into a linear vibration mode.
From Fig.~\ref{Fig3_motion}a-d, it is difficult to directly visualize the noncollinearity for the real and imaginary components, because the angle difference between these two components is very small. Nevertheless, the calculated atom-resolved phonon angular momentum  captures the phonon angular momentum for the Cr and O atoms separately, as shown in Fig.~\ref{Fig3_motion}e,f. 
For Cr atoms located at Wyckoff position $12c$, they have angular momentum along the $z$ axis in the form (0, 0, $L_z$), (0, 0, $-L_z$). For O atoms located at Wyckoff position $18e$, they have in-plane angular momentum \{$I$, $C_3$, $C^2_3$\}($L_x$, 0, 0) that must respect $C_3$ symmetry.
The atom-resolved PAM for both Cr and O are nonzero; however, their sum for \textcolor{black}{any non-degenerate mode, or when summed over a degenerate pair, i.e., the total PAM}, must be zero because $PT$ (parity followed by time reversal) symmetry is preserved for Cr$_2$O$_3$.

\textbf{Discussions}\label{Discussions}

First, the phonon mode frequencies are calculated
in a framework of treating both lattice and spins on equal footing as `slow' degrees of freedom in the current study~\cite{Ren_2024_PhysRevX}. They agree better with the current experimental observations than previous DFT phonon calculations, especially for the $A^1_1$ and $E^1$ modes at around 290\,\cm-1 (see Table~\ref{phonon}). 
\textcolor{black}{At 308\,K, both of these two modes ($A^1_1$ and $E^1$ modes) show Fano asymmetric line shape and are located at different positions (Fig.~\ref{Fig1_RR_RL}a,b). At 18\,K in the AFM phase, the asymmetric lineshape is gone, and they become Lorentzian lineshapes. Surprisingly, they have similar peak frequencies at 18\,K and their frequency difference cannot be resolved by the current experimental resolution of 1.5\,\cm-1 (Fig.~\ref{Fig1_RR_RL}a,b).}
In the phonon frequency calculations, the difference between these two modes is 1\,\cm-1(see Table~\ref{phonon}), much smaller than previous DFT phonon calculations shown in Refs.~\cite{LARBI_2017_Journal,Fechner_2018_PhysRevMaterials}.

The above-mentioned framework of treating both lattice and spins on equal footing as `slow' degrees of freedom is also required to properly treat the splitting of the chiral phonon pairs and spin-phonon coupling. The conventional phonon calculation does not fully capture the impact of broken time-reversal symmetry in magnetically ordered systems. By incorporating the velocity-dependent interatomic forces into the equations of motion and the coupled spin with lattice degrees of freedom, it accounts for time reversal symmetry breaking. 
This framework accurately describes the energies and characteristics of mixed excitations, such as energy splittings between chiral phonon pairs at the Brillouin zone center in the case of CrI$_3$~\cite{Bonini_2023_PRL}. For the case of Cr$_2$O$_3$, there is no energy splitting between the left and right circularly polarized components for the $E$ modes because they are degenerate at the Brillouin zone center. However, their nonzero atom-resolved PAM are correctly captured in this framework~\cite{Ren_2024_PhysRevX}. 

Second, the $A_2$ mode at 266\,\cm-1 appears only at low temperatures, roughly at the same temperature when the defect-induced modes ($E_B^1$ and $E_B^2$) appear.
\textcolor{black}{In the AFM phase with the $D_{3}$ point group symmetries, the existence of Cr vacancies~\cite{Wu_arxiv2025_1} locally breaks the $C_2$ rotational symmetry perpendicular to the three-fold axis, lowering the symmetry to the $C_3$ point group surrounding the Cr vacancies. As a consequence, the $A_2$ antisymmetric Raman scattering tensor gains some symmetric character only when the Cr vacancies become active at low temperatures~\cite{Wu_arxiv2025_1}. This explains why the $A_2$ mode, as well as the $E_u$ modes, only appear at low temperatures.}
Nevertheless, the polarization dependence for the 266\,\cm-1 mode  [Fig.~\ref{Fig1_RR_RL}c] shows that the intensity in the $XX$ scattering geometry is quite small compared with that in the $XY$ scattering geometry, suggesting that the symmetric component is rather small compared with the antisymmetric component.

The local symmetry breaking and the reduction of the point-group symmetry from $D_3$ to $C_3$ around the Cr vacancies may explain why we observe only one $A_{2g}$($A_2$) mode in Cr$_2$O$_3$. The efficiency of intensity transferred from $A_1$ to $A_2$ (to merge into $A$ symmetry due to the local symmetry breaking) depends on the frequency difference between the $A_1$ and $A_2$ modes.
From the DFT phonon calculations shown in Table~\ref{phonon}, 
there are three $A_{2g}$ modes in the nonmagnetic phase: 269, 460, and 674\,\cm-1. 
There are two $A_{1g}$ modes at 293 and 547\,\cm-1. 
The $A_{2g}$ mode at 269\,\cm-1 ($A^1_\text{B}$ mode) has a frequency difference of 24\,\cm-1 with the  $A_{1g}$ mode at 293\,\cm-1.
The $A_{2g}$ mode at 460\,\cm-1 has a frequency difference of 87\,\cm-1 with the  $A_{1g}$ mode at 547\,\cm-1.
The $A_{2g}$ mode at 674\,\cm-1 has a frequency difference of 127\,\cm-1 with the  $A_{1g}$ mode at 547\,\cm-1.
This may explain why we detect only one $A_{2g}/A_{2}$ mode at 266\,\cm-1, as this mode is much closer to a nearby $A_{1g}/A_1$ mode than the other two $A_{2g}/A_{2}$ modes.

Finally, we note that the $A_2$ or $A_{2g}$ phonon has an intrinsic PAM characterized by circular or elliptical motion of atoms. While the total PAM is zero for Cr$_2$O$_3$ due to the $PT$ symmetry, the atom-resolved PAM for the $A_2$ phonon is nonzero and is on the order of $10^{-4}\hbar$, similar to \textcolor{black}{the atom-resolved PAM obtained by summing over the doubly degenerate $E$-mode pairs in Cr$_2$O$_3$.} 
Generally, to have a nonzero PAM for the $A_2$ or $A_{2g}$ phonon mode, inversion symmetry, or time-reversal symmetry, or both need to be broken to lose $PT$ symmetry~\cite{Coh_2023_PhysRevB,Zhang_2026_PRB}. 
The $A_2$ or $A_{2g}$ phonon at an arbitrary $k$ point in a system without $PT$ symmetry can be regarded as an intrinsic chiral phonon. This type of chiral phonon is nondegenerate and does not split in the magnetic field. This differs from the double-degenerate modes at high-symmetry points, especially close to the Brillouin zone center, which split into right- and left-circularly polarized modes in an external magnetic field~\cite{Ahrens_1979_PhysRevLett,Cheng_2020_NanoLetter,Baydin_2022_PhysRevLett,Lujan_2024_PNAS,Mustafa_2025_ACS_Nano,wang_2025_Huang} or below an internal magnetization temperature~\cite{Che_2025_PhysRevLett,Yang_2025_PhysRevLett,Wu_2025_FMO_PRL,Bao_2026_2026}.
While the $A_2$ phonon detected in Cr$_2$O$_3$ cannot be regarded as a chiral phonon, it motivates the search for such an intrinsic chiral phonon at finite momentum in a system without $PT$ symmetry.
    

                                                                                                                                                                                                                                                                                                                                                                                                                                                                                                           
In summary, we detect an $A_2$ phonon at 266\,\cm-1 in the AFM chiral phase of Cr$_2$O$_3$. 
This silent $A_2$ mode becomes Raman-active in Cr$_2$O$_3$ because a minute concentration of Cr vacancies resides on the high-symmetry axis induces a local breakdown of the twofold rotational symmetry perpendicular to the threefold axis.
Our results provide a new strategy to detect this distinct symmetry-forbidden $A_2$ phonon with an antisymmetric Raman tensor in a pseudovector symmetry channel. 
These findings motivate future studies involving this type of $A_2$ mode; for instance, exciting such phononic excitations on an ultrafast timescale may offer a pathway to induce geometric chirality in otherwise achiral crystals~\cite{Romao_2024_acsnano}. 
                                 
\bigskip

\textbf{Methods}

\textbf{Single crystal preparation and characterization.}\label{Crystal_preparation}                                                                                                    
Single crystals of Cr$_2$O$_3$ were synthesized by a laser floating zone technique at Rutgers university described in Ref.~\cite{Du_2023_NPJ}. The Cr$_2$O$_3$ sample was characterized by magnetic susceptibility measurements in a warming-up process from 5\,K to 370\,K with magnetic field $H=100$\,Oe along the hexagonal $c$ axis after the zero magnetic field cooling (ZFC) to 5\,K.
The extracted antiferromagnetic order transition temperature is $T_\text{N}=307.5$\,K~\cite{Wu_arxiv2025_1}.
The sharp Raman modes and the low residual spectra background indicate the high quality of the single crystals [Fig.~\ref{Fig1_RR_RL} of main text]. 
\textcolor{black}{
In our previous study~\cite{Wu_arxiv2025_1}, we have shown that the Cr$_2$O$_3$ sample contains some dilute Cr vacancies.
}

\textbf{Raman scattering measurements.}\label{Raman}                                        
The Cr$_2$O$_3$ samples were polished in the air to expose a (0~0~1) hexagonal crystallographic $ab$ plane.  
A strain-free area was examined by a Nomarski image. The polished crystals were positioned in
a continuous helium flow optical cryostat. The Raman measurements
were mainly performed using the Kr$^+$ laser line at 647.1\,nm (1.92\,eV) in
a quasibackscattering geometry along the crystallographic $c$ axis.
The excitation laser beam was focused into a $50\times100$ $\mu$m$^2$
spot on the $ab$ surface, with the incident power around 10\,mW. The
scattered light was collected and analyzed by a triple-stage Raman
spectrometer, and recorded using a liquid-nitrogen-cooled
charge-coupled detector. 
Linear and circular polarizations were used in this study to decompose the Raman data into different irreducible representations.
The instrumental resolution was maintained better than 1.5\,\cm-1. 
The temperatures were corrected for laser heating (see Supplementary Note~\ref{laser_heating_determination}). 
                                                                                                                                                                                                                                                             
All spectra shown were corrected for the spectral response of the spectrometer and charge-coupled detector to obtain the Raman intensity $I
_{\mu v}$, which is related to the Raman response $\chi''(\omega,T)$: $I_{\mu v}(\omega, T)=[1+n(\omega, T)] \chi_{\mu \nu}^{\prime \prime}(\omega, T)$. Here $\mu (v)$ denotes the polarization of the 
incident (scattered) photon, $\omega$ is the energy, $T$ is the temperature, and $ n(\omega, T)$ is the Bose factor.
                                                                                                                                                                                        
The Raman spectra were recorded from the $ab$ (0~0~1) surface for scattering geometries denoted as $\mu v = XX, XY, RR, RL$, which is short for $Z(\mu v)\bar{Z}$ in Porto’s notation, where $X$ and $Y$ denote linear polarization parallel and perpendicular to the crystallographic axis, respectively; $R=X+iY$ and $L=X-iY$ denote the right- and left-circular polarizations, respectively. The $Z$ direction corresponds to the $c$ axis perpendicular to the (0~0~1) plane. 
The polarization leakage from optical elements was removed in our data analysis (see Supplementary Note~\ref{leakage}).

 

\textbf{Density functional theory calculations.}\label{DFT}
Density functional theory calculations were performed using the Vienna \textit{ab initio} simulation package~\cite{Kresse1996_PhysRevB,KRESSE199615}, employing the local-density-approximation (LDA) exchange-correlation functional~\cite{Perdew_1981_PhysRevB} and the projector-augmented-wave~\cite{Blochl_1994_PhysRevB} method, with Cr~3s$^2$3p$^6$3d$^5$4s$^1$, O~2s$^2$2p$^4$ pseudo potential valence configurations. A plane-wave cutoff of 500\,eV is chosen for the Cr$_2$O$_3$ system. 
The structure is relaxed using the local spin-density approximation, and the convergence criteria for forces and energies are $10^{-3}$\,eV/\AA~and  $10^{-8}$\,eV, respectively. After relaxation, the wave functions are calculated using static calculations with a convergence criterion of 10$^{-10}$\,eV for energies. Spin-orbit coupling is included in all static calculations except structural relaxations.
The $\Gamma$-centered Monkhorst-Pack $k$-points meshes with $7 \times 7 \times 7$ is used for the calculation of Cr$_2$O$_3$. The Dudarev-type DFT + U approach~\cite{Dudarev_1998_PhysRevB} is used with values of $U=4.0$\,eV and $J=0.6$\,eV for Cr atoms.
To ensure consistency with the experimental ground state, a 2\% epitaxial strain is applied in our calculations, as our approach assumes spins to be oriented along the $z$ direction~\cite{Mu_2019_PhysRevMaterials}.

\textbf{Symmetry analysis.}\label{space_group}
The information for the irreducible representations of point groups and space groups follows the notations of Cracknell~$et~al$~\cite{cracknell1979general}, which is the same for the Bilbao Crystallographic Server~\cite{Bilbao_4,Bilbao_2}.

\textbf{Data availability}

The data generated in this study are available from the corresponding authors upon request.

\bibliographystyle{naturemag}

 \newpage   

\textbf{Acknowledgments}

The work at BAQIS (S.W.) was supported by the National Natural Science Foundation of China (Grant No.~12404548).  
The spectroscopy work at Rutgers (S.W. and G.B.) was supported by NSF Grants No.~DMR-2105001. 
The sample growth and characterization (K.D., X.X., and S.-W.C.) was supported by the W. M. Keck foundation grant to the Keck Center for Quantum Magnetism at Rutgers University.
The DFT work at Rutgers (S.R. and D.V.) was supported by NSF Grants No.~DMR-2421895. 
The work at NICPB was supported by the European Research Council (ERC) under the European Union’s Horizon 2020 research and innovation programme grant agreement No.~885413.

\textbf{Author Contributions} 

G.B. supervised the experiments. 
S.W. and G.B. acquired and analyzed the Raman spectra. 
S.R. and D.V. performed the DFT phonon calculations. 
K.D., X.X., and S.C. grew and polished the single crystals. 
All the authors contributed to the discussions and writing of the paper.

$^\star$ S.W. and S.R. contributed equally to this work.
    
\textbf{Competing interests} 

The authors declare no competing interests.

\textbf{Materials \& Correspondence}

Correspondence and requests for materials should be addressed to Shangfei~Wu and Girsh~Blumberg.

\newpage

%
\renewcommand{\thefigure}{S\arabic{figure}}
\addtocounter{figure}{-3}
\renewcommand{\theequation}{S\arabic{equation}}
\addtocounter{equation}{-0}

\textbf{Supplementary Materials}

\section{Group-theoretical analysis}\label{GroupTheory}

Above $T_\text{N}$, Cr$_2$O$_3$ belongs to the rhombohedral structure with space group $R\bar{3}c$ (No.~167) [point group $D_{3d}$]. The Cr and O atoms have Wyckoff positions $12c$ and $18e$, respectively.                                                             
From the group theoretical considerations, $\Gamma$ point phonon modes of the Cr$_2$O$_3$ can be expressed as $\Gamma$ = 2$A_{1g}$ $\oplus$ 2$A_{1u}$ $\oplus$ 3$A_{2g}$ $\oplus$ 3$A_{2u}$ $\oplus$ 5$E_{u}$ $\oplus$ 5$E_{g}$. Raman active modes $\Gamma_{\text{Raman}}$= 2$A_{1g}$ $\oplus$ 5$E_g$, infrared active modes are $\Gamma_{\text{IR}}$ = 2$A_{2u}$ $\oplus$ 4$E_{u}$, and acoustic modes $\Gamma_{\text{acoustic}}$ = $A_{2u}$ $\oplus$ $E_{u}$. \textcolor{black}{Note that $A_{2g}$ is neither Raman nor infrared active. It might become Raman active under resonant conditions~\cite{Placzek1934}.}

For the $D_{3d}$ point group, the Raman selection rules indicate that the $XX$, $XY$, $RR$, and $RL$ polarization geometries probe the $A_{1g} + E_g$, $A_{2g}+ E_g$, $A_{1g} + A_{2g}$, and $2E_g$ symmetry excitations, respectively. The relationship between the scattering geometries and the probed symmetry channels are summarized in Table~\ref{SymmetryAnalysis}. 
                                            
\begin{table}[b]
\caption{\label{SymmetryAnalysis}The relationship between the scattering geometries and the symmetry channels for point groups $D_{3d}$ and $D_3$. $A_{1g}$,  $A_{2g}$, and $E_{g}$ are the irreducible representations of the $D_{3d}$ point group, while $A_{1}$,  $A_{2}$, and $E$ are the irreducible representations of the $D_{3}$ point group.}
\begin{ruledtabular}
\begin{tabular}{ccc}
Scattering geometry&$D_{3d}$& $D_{3}$\\
\hline
$XX$&$A_{1g}+E_{g}$&$A_{1}+E_{}$\\
$XY$&$A_{2g}+E_{g}$&$A_{2}+E_{}$\\
$RR$&$A_{1g}+A_{2g}$&$A_{1}+A_{2}$\\
$RL$&$2E_{g}$&$2E$\\
\end{tabular}
\end{ruledtabular}
\end{table}
                                       

Below $T_\text{N}$, the presence of magnetic ordering in bulk Cr$_2$O$_3$  breaks $i$, $\sigma_d$, and $S_6$ symmetries associated with the $D_{3d}$ point group symmetries. Consequently, the original $D_{3d}$ point group symmetry is reduced to $D_3$ below $T_\text{N}$, resulting in interirrep mixing between bare phonons of different irreducible representations (irreps). The $E_g$ and $E_u$ irreps become the $E$ irrep, the $A_{1g}$ and $A_{1u}$ become the $A_1$ irrep, and the $A_{2g}$ and $A_{2u}$ irreps become the $A_2$ irrep of the $D_3$ point group [Table~\ref{decompositionD3dToD3}].
\textcolor{black}{From the group theoretical considerations, $\Gamma$ point optical phonon modes of the Cr$_2$O$_3$ can be expressed as $\Gamma_{\text{optical}}$ = 4$A_1$ + 6$A_2$ + 10$E$ and the acoustic modes are $\Gamma_{\text{acoustic}}$ =$A_2$ $\oplus$ $E$. 
All the $E$-symmetry optical modes are both Raman and infrared active. The $A_1$-symmetry modes are Raman active, while the $A_2$-symmetry modes turn into infrared active.}


\begin{table}[b]
\caption{\label{decompositionD3dToD3} The relationship of the irreducible representations between $D_{3d}$ and $D_3$ point groups.}
\begin{ruledtabular}
\begin{tabular}{cc}
$D_{3d}$&$D_3$\\
\hline
$A_{1g}$&$A_1$\\
$A_{1u}$&$A_1$\\
$A_{2g}$&$A_2$\\
$A_{2u}$&$A_2$\\
$E_{g}$&$E$\\
$E_{u}$&$E$\\
\end{tabular}
\end{ruledtabular}
\end{table}   

%

\section{Laser heating determination} \label{laser_heating_determination}                                                                                                                             
The laser heating rate, a measure of the temperature increase per unit laser power in the focused laser spot (K/mW), in the Raman experiments was determined by monitoring 
the appearance of one-magnon scattering peak during the laser heating process at room temperature 296\,K. 
With laser power 10\,mW, there is a clear one-magnon scattering peak at  3\,\cm-1, indicating the laser spot temperature is below $T_\text{N}$=307.5\,K.
When heating with laser power of 15\,mW, the one-magnon mode weakens significantly and shifts to lower energy 2.8\,\cm-1, indicating the laser spot temperature is slightly below $T_\text{N}$=307.5\,K.
When heating with laser power of 25\,mW, the one-magnon mode disappears, indicating the laser spot temperature is above $T_\text{N}$=307.5\,K. Thus, the heating coefficient can be determined via the constraint: $296\,\text{K}+15\,\text{mW}*k  \approx 307.5$\,K. In this way, we have deduced the heating coefficient: $k \approx 0.77\pm0.1$\,K/mW.

\section{Removal of polarization leakage}\label{leakage}

In this section, we provide the detailed procedure to remove the polarization leakage signal from optical elements in our data analysis. 

In our polarization optics setup, we used a Glan-Taylor polarizing prism (Melles Griot) with an extinction ratio better than $10^{-5}$ to clean the laser excitation beam and a broadband 50\,mm polarizing cube (Karl Lambrecht Corporation) with an extinction ratio better than 1:500 to analyze the scattered light.
For the linearly polarized $XX$ and $XY$ scattering geometries, the leakage intensity ratio is negligibly small (less than 0.2\,\%); thus, the leakage intensity from the linearly polarized scattering geometries is not considered.
     
For measurements with circularly polarized light, we employed a Berek compensator (New Focus) to convert the incoming linearly polarized light into circularly polarized light for excitation. We used a broadband 50-mm-diameter quarter-wave retarder (Melles Griot) with a retardance tolerance $\lambda/50$ before the polarizing cube to convert the outgoing circularly polarized light into linearly polarized light for the analyzer. 
The leakage of circular polarized light is due to the limitations of the broadband quarter-wave plate and
alignment of the Berek compensator.  

\begin{figure}[!t] 
\begin{center}
\includegraphics[width=0.8\columnwidth]{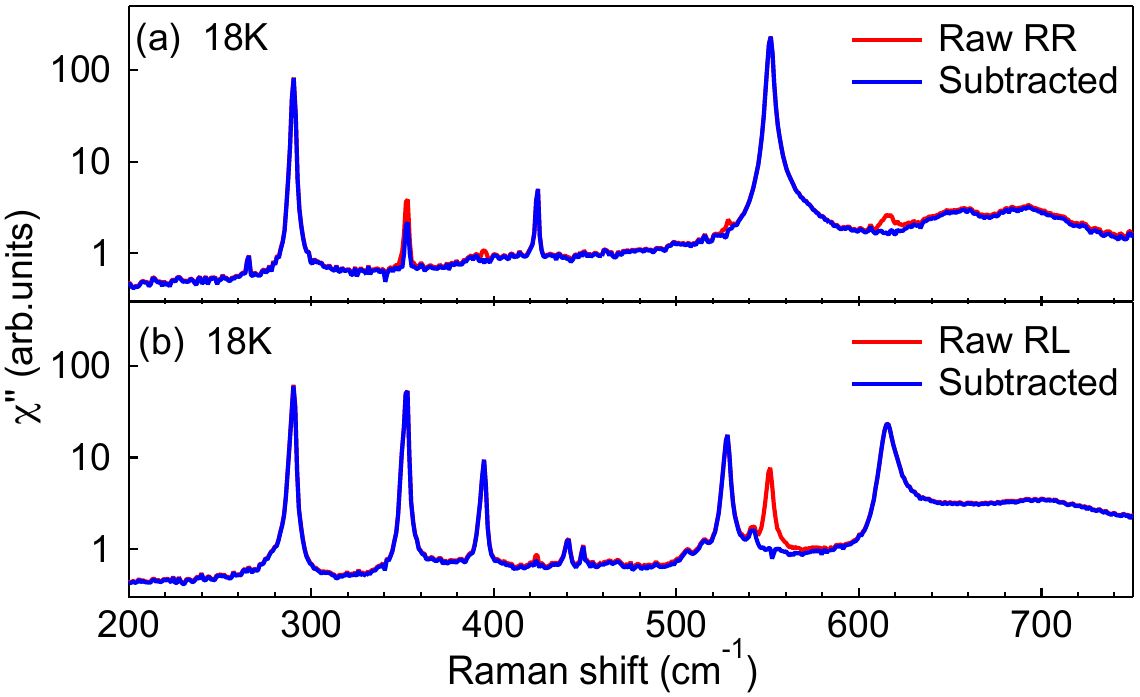}
\end{center}
\caption{\label{Fig_S2_RM_leakage} Comparison of raw data and polarization leakage removed spectra, taken in RR (panel \textbf{a}) and RL (panel \textbf{b}) polarization scattering geometry from the $ab$ surface of Cr$_2$O$_3$ at 18\,K with 647-nm laser excitation.}
\end{figure}    
                                                                                                           
The amount of phonon leakage intensity is determined based on the
$A_{1}$  and $E$ bulk phonons of Cr$_2$O$_3$ at the same temperature in $RR$ and $RL$ scattering geometries. 
We follow the procedures presented in Ref.~\cite{Wu_2022_PhysRevB} to remove the polarization leakage intensity.                                                                                                                                                                                                                              In Fig.~\ref{Fig_S2_RM_leakage}, we show the Raman spectra of the unprocessed raw data and
polarization-leakage-removed spectra taken at 18\,K from the
$ab$ surface of Cr$_2$O$_3$ crystals in $RR$ and $RL$ scattering geometries,
respectively. The leakage intensity of the bulk $E$ and $A_{1}$ 
phonons in the raw data can be fully removed with a leakage ratio close to 2\%, except for the mode at about 350\,\cm-1. 

The 350\,\cm-1 mode is an $E$ mode in the $RL$ scattering geometry with a leakage ratio of about 6\% into the $RR$ scattering geometry. The leakage intensity in the $RR$ scattering geometry cannot be fully removed with an average leakage ratio of 2\% using the procedures above~\cite{Wu_2022_PhysRevB}. The origin of the anomalous leakage intensity for the 350\,\cm-1 mode is unknown. It might be related to the polarization rotations within the AFM phase~\cite{Krichevtsov_1993_JOP}, as the leakage intensity in the $RR$ scattering geometry only appears below $T_\text{N}$~[Fig.~\ref{Fig1_RR_RL} of main text].

\end{document}